# The only non-contradictory model of universe

## Vladimír Skalský

Faculty of Materials Science and Technology of the Slovak Technical University, 917 24 Trnava, Slovakia, skalsky@mtf.stuba.sk

**Abstract.** The Friedmann equations of universe dynamics describe the infinite number of the Friedmannian models of universe. The consistent and distinguished relativistic, classical-mechanical, quantum-mechanical and formal-logical analysis of the Friedmannian universe models leads to a surprising and unexpected conclusion: The Friedmannian model of the flat expansive homogeneous and isotropic universe with the zero gravitational force state equation is the only model of universe, which does not contradict the: 1st Einstein general theory of relativity (and its special partial solutions: the Einstein special theory of relativity and the Newton theory of gravitation); 2nd quantum mechanics; 3rd fundamental formal principles of logical thinking; and 4th observations.

## 1. Introduction

The mathematical-physical basis of the present relativistic cosmology represents *the Friedmann general equations of isotropic and homogeneous universe dynamics* (Friedmann 1922, 1924). Using *the Robertson-Walker general metrics of isotropic and homogeneous universe* (Robertson 1935, 1936a, b; Walker 1936) they can be written in the form:

$$\dot{a}^2 = \frac{8\pi G a^2 \rho}{3} - kc^2 + \frac{\Lambda a^2 c^2}{3}, \tag{1a}$$

$$2a\ddot{a} + \dot{a}^2 = -\frac{8\pi G a^2 p}{c^2} - kc^2 + \Lambda a^2 c^2, \tag{1b}$$

$$p = K\varepsilon, \tag{1c}$$

where $a$ is the gauge factor, $\rho$ is the mass density, $k$ is the curvature index, $\Lambda$ is the cosmological member, $p$ is the pressure, $K$ is the state equation constant and $\varepsilon$ is the energy density.

The Friedmannian equations (1a), (1b) and (1c) – without introducing of any restrictive supplementary assumptions – describe infinite number of *the Friedmannian models of universe.* Individual Friedmannian models of universe are determined by the equations (1a), (1b) and (1c) with the values: $k = +1, 0, -1$; $\Lambda > 0, = 0, < 0$; and $K > 0, = 0, < 0$.

The Friedmannian equations (1a), (1b) and (1c) represents an application of *the Einstein field equations* (Einstein 1915) to the whole isotropic relativistic universe in a Newtonian approximation, under the hypothetical supplementary assumptions, introduced into the relativistic cosmology by Einstein (1917), de Sitter (1917) and Friedman (1922, 1924), generalised by Friedmann (1922, 1924) and transparent by Robertson (1935, 1936a, b) and Walker (1936).

The Friedmannian equation (1a), (1b) and (1c) contain the supplementary assumptions in the mathematically generalised form, therefore, they contains and its factually negation, too. This means that they describe the Friedmannian models of universe which are solution of the Einstein field equations in the Newtonian approximation with the hypothetical supplementary assumptions (introduced into the relativistic cosmology by Einstein, de Sitter and Friedmann) and without they, too (i.e. with all possible combinations its non-zero and zero values).

Therefore – under assumption that the mathematical-physical description of our observed *expansive and isotropic relativistic Universe* not require an introduction of any next supplementary assumption – the Friedmannian equations (1a), (1b) and (1c) must immanently contained and the Friedmannian model of universe, which described our observed Universe, too (Skalský 1997).

## 2. Observed properties of the Universe

According to the observations, we reliably know that:

**The observed Universe is a) *relativistic*, b) *expansive*, and c) *isotropic*.**

The relativisticity of the Universe is unambiguously confirmed by quite a number of observations of the relativistic effects on the Earth, in its vicinity, in the vicinity of the Sun, in the Solar System and in the further Universe, too.

E. P. Hubble discovered the expansion of the Universe in 1929 (Hubble 1929).

The observations unambiguously demonstrate that in the Universe in the largest distances in all directions there exist equal numbers of the cosmic objects with equal properties.

In the cosmological literature is stated that:

**$d_1$) The observed expansive and isotropic relativistic Universe (at larger distances) is *homogeneous*, too.**

However, this statement contradicts the observations and the general theory of relativity, too! The observed expansive and isotropic relativistic Universe principally cannot be homogeneous! In the expansive and isotropic relativistic Universe the signal propagate at finite velocities, therefore, if we look into the distance we observe the events (i.e. from the relativistic point of view we are their contemporaries), which (from the Newtonian point of view) were realised in ancient past. In larger distances we observe the Universe with smaller dimensions, bigger mass density, higher temperature … and with such processes, which we do not observe in our near Universe surrounding. The fact that they are not some small, irrelevant differences we can demonstrate for example on the mass density of our Universe $\rho$. According to the present cosmological literature, is the estimated value of present mass density of Universe $\rho_{pres} \sim 10^{-26}$ to $10^{-32}$ kg m$^{-3}$. According to *the Planck quantum hypothesis*, the Universe began its expansive evolution with *the Planck mass density* $\rho_{Planck} \sim 10^{97}$ kg m$^{-3}$. This means that difference of the mass density of Universe $\rho$ in our nearest Universe surrounding and in the furthest Universe is 123 to 129 ranks!

Therefore, we can conclude:

**$d_2$) The observed expansive and isotropic relativistic Universe is *inhomogeneous*.**

The observed expansive and isotropic relativistic Universe can be regarded as homogeneous only in approximation in which we abstract from the relativistic effects, i.e. only in *the first (Newtonian) approximation*!

Therefore:

**$d_3$) The observed expansive and isotropic relativistic Universe can be regarded (at larger distances) as homogeneous only in the Newtonian approximation.**

However, between the thesis $d_3$ and the thesis $d_1$ is a substantial and principal difference!

The substitution of the thesis $d_3$ by the thesis $d_1$ is one of serious mistakes, which are the components of *the present cosmological paradigm*.

## 3. The model properties of the expansive and isotropic relativistic Universe

In the observed expansive and isotropic relativistic Universe the gravitation and expansion of matter objects cause the relativistic effects, which for the observers in any co-ordinate systems have different values and in the largest distances, in which the velocity of its expansion approximates to *the boundary velocity of signal propagation*, they gain extremely values, which approximate to the limit (i.e. zero, or infinite) values.

Therefore:

**1. The global parameters of the observed expansive and isotropic relativistic Universe (with non-limit values) principally cannot be possibly expressed relativistically!**

We can express them only in the non-relativistic approximation, i.e. only using such a theory of gravitation in which we abstract from the observed relativistic effects.

There exists only one theory of gravitation:

- *the Einstein general theory of relativity* (Einstein 1915, 1916),

   which exactly describes the macro-physical reality, and has only one special partial solution in which we abstract from all relativistic effects:

- *the Newton theory of gravitation (the classical mechanics)* (Newton 1687).

The general theory of relativity is the macro-physical (relativistic) theory of (inertial and non-inertial) co-ordinate systems, which describes the relativistic properties of the physical objects from point of view of any co-ordinate system; the global relativistic point of view does not exist. Therefore, from point of the general theory of relativity the concepts *whole of universe* and *whole parameters of universe* have not a concrete physical sense!

From point of view of the general theory of relativity about 'the whole of Universe' and about 'the whole parameters of Universe' we can consider only limitly. However, the limit relativistic parameters of the expansive and isotropic relativistic Universe have non-physical (i.e. zero, or infinite) values.

We cannot describe the whole of the expansive and isotropic relativistic Universe and its global parameters in a relativistic non-limit way just because the general theory of relativity precisely describes the macro-physical (relativistic) reality.

What it is not possible in the general theory of relativity (because it precisely describes all parts of relativistic reality), becomes trivial in the Newton theory of gravitation. It is because in the Newton theory of gravitation we abstract just from the relativistic effects which in the general theory of relativity such a non-limit description make impossible.

Therefore, we can state:

**2. The global parameters of the expansive and isotropic relativistic Universe with finite (non-limit) values of the physical quantities can be expressed only in the Newtonian approximation.**

The whole parameters of the expansive and isotropic relativistic Universe in the Newtonian approximation (projection) are equivalent to the parameters of the Newtonian-Euclidean expansive homogeneous matter sphere.

This fact has a significant influence on the model properties of observed expansive and isotropic relativistic Universe, because:

**3. The Euclidicity of space in the Newton theory of gravitation principally excludes any consideration on curvature of space in the Newtonian approximation of the expansive and isotropic relativistic Universe!**

The shown facts have these results:

**4a. The expansive and isotropic relativistic Universe in the Newtonian model projection (approximation) is flat (Euclidean), irrespective of actual relativistic matter-space-time properties it has.**

**4b. From the fact 4a reciprocally results that the actual observed expansive and isotropic relativistic Universe principally is not, and cannot be, flat.**

The fact that the global parameters of the expansive and isotropic relativistic Universe can be non-limitly expressed only in the Newtonian model projection, however, it does not mean that the Universe is Newtonian. The observed Universe is relativistic, it means that the Universe in the Newtonian model approximation must consider these relativistic facts:

As a consequence of the finite velocity of signal propagation we can *optically* observe the Universe up to the beginning of matter era, when – as a result of *"the recombination" of hydrogen atoms* – the matter and the radiation were separated and the Universe became transparent for the photons. Using *the neutrino observing technology* we could see even further, and using *the hypothetical graviton observing technology* we could see up to the cosmological time $t \sim 10^{-43}$ s. Theoretically (using the retrospective extrapolation), we can "reach" even into the beginning of Universe expansive evolution, i.e. up to *the initial limit cosmological singularity*. This means that we are contemporaries of the expansive evolution of Universe at any physically real cosmological times $t$ and theoretically (in the mathematical-physical sense) we are also contemporaries of the initial limit cosmological singularity.

According to Einstein, in the finite area no choice of co-ordinates can exclude the gravitational field, however, infinitely small area of space can be considered as flat, in which the laws of the special theory of relativity are valid (Einstein 1921). Therefore, the theoretical initial limit cosmological singularity with zero dimensions – from the point of view of the general theory of relativity – behaves in two ways:

- As a gravitational object with the limit matter-space-time values in which the motion stopped.

- As an inertial system with the limit matter-space-time values, which expands towards all observers at the boundary velocity of signal propagation.

These facts in the Newtonian model approximation are considered only by the model of Newtonian-Euclidean homogeneous matter sphere, expanding at the constant velocity $v = c$.

This means:

**5. The expansive and isotropic relativistic Universe in the Newtonian model approximation during the whole expansive evolution expands at the boundary velocity of signal propagation $c$.**

Therefore, for the gauge factor of universe $a$, i.e. for the radius of Euclidean homogeneous matter sphere $r$ and the cosmological time $t$ in the Newtonian model approximation of the expansive and isotropic relativistic Universe are valid the relations (Skalský 1992, 1989):

$$a = r = ct \,. \tag{2}$$

The Newtonian-Euclidean homogeneous matter sphere expanding at the velocity $v = c$ has the radius

$$r = r_g = \frac{2Gm}{c^2}, \tag{3}$$

where $r_g$ is the Schvarzschild gravitational radius and $m$ is the mass.

6. **In the Newtonian model approximation of the expansive and isotropic relativistic Universe – represented by the Newtonian-Euclidean homogeneous matter sphere, expanding at the constant velocity $v = c$ – the gravitational forces cannot manifest themselves.**

According to the general theory of relativity, *the gravitational forces* are determined by the sum of the energy density $\varepsilon$ and three-multiple of the pressure $p$. Therefore, for the sum of the energy density $\varepsilon$ and three-multiple of the pressure $p$ in the Newtonian model approximation of the expansive and isotropic relativistic Universe is valid the relation (Skalský 1991):

$$\varepsilon + 3p = 0 \,. \tag{4}$$

The general theory of relativity and *the quantum mechanics* are the complementary theories. The macro-physical relativistic universe is in finite result represented by the microphysical quantum-mechanical objects (by the particles and the fields) and on the contrary, the quantum-mechanical objects can really exist only in the relativistic macro-world (universe). This means that the Newtonian model of the expansive and isotropic relativistic Universe, which fulfil the restrictive conditions, resulting from the general theory of relativity in the Newtonian approximation, simultaneously must also fulfil the restrictive conditions, which result from the quantum mechanics *(the quantum theory)*.

According to *the Planck quantum hypothesis (theory)*, it has a sense to think about the physical parameters of the observed expansive Universe from the moment when its dimensions reach the values which correspond to *the Planck length* $l_P = (hG/c^3)^{1/2}$, i.e. at *the Planck time* $t_P = (hG/c^5)^{1/2}$ (Planck 1899).

At present time *the Planck units* are shown with *the Planck (reduced) constant* $\hbar = h/2\pi$:

*the Planck length* (Cohen and Taylor 1999)

$$l_P = \sqrt{\frac{\hbar G}{c^3}} = ct_P = 1.616\,05(10) \times 10^{-35}\,\text{m}, \tag{5}$$

*the Planck time* (Cohen and Taylor 1999)

$$t_P = \sqrt{\frac{\hbar G}{c^5}} = \frac{l_P}{c} = 5.390\,56(34) \times 10^{-44}\,\text{s} \,. \tag{6}$$

From definitions of the relations (5) and (6) it results unambiguously that:

7. **According to the Planck quantum theory, the expansive universe could begin its expansive evolution at only one possible velocity $v = c$, and at the cosmological time $t = t_P$ (6) for the gauge factor $a = l_P$ (5) the relations (2) were valid in it.**

We can conclude:

8. **From infinite numbers of the Friedmannian models of the flat expansive isotropic and homogeneous universe – determined by the Friedmannian equations (1a), (1b) and (1c) with $k = 0$, $\Lambda = 0$ and $K > 0, = 0, < 0$ – only one model (i.e. only with one from the values of the state equation constant $K$) can fulfil *the relativistic and quantum-mechanical restrictive conditions*, expressed by the relations (2), (3), (4), (5) and (6).**

## 4. The Friedmannian model of the expansive and isotropic relativistic universe

The retrospective extrapolation of the evolution of expansive and isotropic relativistic universe leads to the theoretical initial limit cosmological singularity, therefore, in the Newtonian model approximation of the universe could begin its expansive evolution only at one possible velocity: the boundary velocity of signal propagation $c$.

According to the special theory of relativity, with the boundary velocity of signal propagation $c$ the time-flow will stop, therefore, the expansive and isotropic relativistic universe, which started its expansive evolution at the boundary velocity (at which the time was stopped), in the Newtonian model approximation (at which the time flow equally with an arbitrary velocity) it must during the whole expansive evolution expand at the constant velocity $v = c$ (Skalský 1992, 1989).

In the Newtonian model projection of the expansive and isotropic relativistic universe (i.e. in the Newtonian-Euclidean homogeneous matter sphere, expanding at the constant velocity $v = c$), all parameters are mutually linearly bound, therefore, **the relation between its two arbitrary model parameters can be used as a restrictive condition for the unambiguously determination of the Friedmannian universe model, which describes our observed Universe.**

The Friedmannian equations (1a), (1b) and (1c) with $k = 0$ and $\Lambda = 0$ fulfil the relativistic and quantum-mechanical restrictive condition of the relations (2) (and the relativistic and the quantum-mechanical restrictive conditions of the relations (3), (4), (5) and (6), too) only with the value of state equation constant $K = -1/3$, i.e. only with (Skalský 1991)

*the zero gravitational force state equation*

$$p = -\frac{1}{3}\varepsilon. \tag{7}$$

Using the Friedmannian equations (1a) and (1b) with $k = 0$ and $\Lambda = 0$, the state equation (7), the Newtonian relations for Euclidean homogeneous matter sphere and the Hubble relation (Hubble 1929) we can determine the relation between next parameters of the expansive and isotropic relativistic universe, in the Newtonian approximation, i.e. between other parameters of the Friedmannian model of *the* (flat) *expansive non-decelerative* (isotropic and homogeneous) *universe* (ENU) (Skalský 1991):

$$a = ct = \frac{c}{H} = \frac{2Gm}{c^2} = \sqrt{\frac{3c^2}{8\pi G\rho}}, \tag{8}$$

where $H$ is the Hubble coefficient ("constant").

For better transparency we express the relations (7) and (8) in all next possible relations and variants (Skalský 1997, 2000d):

$$t = \frac{a}{c} = \frac{1}{H} = \frac{2Gm}{c^3} = \sqrt{\frac{3}{8\pi G\rho}}, \tag{9}$$

$$H = \frac{c}{a} = \frac{1}{t} = \frac{c^3}{2Gm} = \sqrt{\frac{8\pi G\rho}{3}}, \tag{10}$$

$$m = \frac{c^2 a}{2G} = \frac{c^3 t}{2G} = \frac{c^3}{2GH} = \sqrt{\frac{3c^6}{32\pi G^3 \rho}}, \tag{11}$$

$$\rho = \frac{3c^2}{8\pi G a^2} = \frac{3}{8\pi G t^2} = \frac{3H^2}{8\pi G} = \frac{3c^6}{32\pi G^3 m^2} = -\frac{3p}{c^2}, \tag{12}$$

$$p = -\frac{c^4}{8\pi G a^2} = -\frac{c^2}{8\pi G t^2} = -\frac{c^2 H^2}{8\pi G} = -\frac{c^8}{32\pi G^3 m^2} = -\frac{c^2 \rho}{3} = -\frac{1}{3}\varepsilon. \tag{13}$$

From the relations (8)–(13) it result the relations for fundamental mass-space-time parameters of ENU and their variants (Skalský 1997):

$$m = Ca = Dt, \tag{14}$$

$$a = ct = C'm, \tag{15}$$

$$t = c'a = D'm, \tag{16}$$

where $C$, $D$, $C'$ and $D'$ are the (total) constants:

$$C = \frac{1}{C'} = \frac{m}{a} = \frac{m}{ct} = \frac{Hm}{c} = \sqrt{\frac{8\pi G \rho m^2}{3c^2}} = \frac{c^2}{2G} = 6.734\,67(15) \times 10^{26}\ \text{kg}\,\text{m}^{-1}\,, \tag{17}$$

$$C' = \frac{1}{C} = \frac{a}{m} = \frac{ct}{m} = \frac{c}{Hm} = \sqrt{\frac{3c^2}{8\pi G \rho m^2}} = \frac{2G}{c^2} = 1.484\,85(34) \times 10^{-27}\ \text{m}\,\text{kg}^{-1}\,, \tag{18}$$

$$D = \frac{1}{D'} = \frac{cm}{a} = \frac{m}{t} = Hm = \sqrt{\frac{8\pi G \rho m^2}{3}} = \frac{c^3}{2G} = 2.019\,00(37) \times 10^{35}\ \text{kg}\,\text{s}^{-1}\,, \tag{19}$$

$$D' = \frac{1}{D} = \frac{a}{cm} = \frac{t}{m} = \frac{1}{Hm} = \sqrt{\frac{3}{8\pi G \rho m^2}} = \frac{2G}{c^3} = 4.952\,93(79) \times 10^{-36}\ \text{s}\,\text{kg}^{-1}\,, \tag{20}$$

and

$$c' = \frac{1}{c} = 3.335\,640\,951\,981\,520\ldots \times 10^{-9}\ \text{s}\,\text{m}^{-1}\,. \tag{21}$$

From the relations (8)-(13) it results the value of deceleration parameter $q$ in the ENU:

$$q = -\frac{a\ddot{a}}{\dot{a}^2} = 0\,. \tag{22}$$

**The flatness (Euclidicity), expansivity, non-decelerativity (zero, positive or negative acceleration), isotropy and homogeneity of the Friedmannian model of the ENU result from the mutual linear dependence of all its parameters – the gauge factor $a$, cosmological time $t$, Hubble coefficient $H$, mass $m$, (critical) mass density $\rho$, or (critical) energy density $\varepsilon$, and (negative) pressure $p$ – in the only possible manner of the relations (8)-(13) at which the gravitational forces cannot manifest themselves.**

We can check it by a simple calculation:

The ENU – according to the relations (8) – at any cosmological time $t$ in the distance of gauge factor $a$ expands at the boundary velocity of signal propagation $c$. Therefore, the radial velocity of ENU expansion $v_r$ in an arbitrary distance $r \leq a$ is given by the relation:

$$v_r = \frac{r}{a} c\,. \tag{23}$$

In the Newton theory of gravitation for the mass of a homogeneous matter sphere $m$ with the radius $r$ and the mass density $\rho$ is valid the relation:

$$m = \frac{4}{3}\pi r^3 \rho\,. \tag{24}$$

The escape (second cosmic) velocity $v_2$ from the surface of the homogeneous matter sphere with the mass $m$ and the radius $r$ in the Newton theory of gravitation is determined by the relation:

$$v_2 = \sqrt{\frac{2Gm}{r}}\,. \tag{25}$$

**From the relations (12), (23), (24) and (25) it results that in the ENU, the homogeneous matter sphere with an arbitrary radius $r \leq a$ and the (critical) mass density of the ENU $\rho$ expands at the velocity $v_r = v_2$.**

(Let's compare the values of the velocities $v_r$ and $v_2$ determined by the relations (23) and (25) with an arbitrary $a$ and corresponding $\rho$, an arbitrary $r \leq a$ and corresponding $m$.)

**This means that the gravitation in the ENU – at the range of its expansivity, nondecelerativity, isotropy and homogeneity – is exactly compensated by its expansion.**

The mentioned fact also explains why the ENU during the whole expansive evolution can expand at the velocity $v = c$, in spite of its non-zero mass density $\rho$.

Therefore, we can state:

**The gravitational interaction in the expansive and isotropic relativistic universe – the global parameters of which we can express only in the Newtonian approximation by the ENU model relations (8)-(13) – represents a local disturbance of symmetry, which is manifested only when the local mass density of matter objects $\rho_{loc}$ is larger than the critical mass density $\rho$.**

This conclusion is in good agreement with the present observations and estimations:

In the present cosmological literature is shown *the present estimated cosmological time of our Universe*

$$t_{pres} \sim (1.2-1.8) \times 10^{10} \text{ yr} \sim (3.8-5.7) \times 10^{17} \text{ s}. \tag{26}$$

If our observed expansive and isotropic relativistic Universe in the Newtonian approximation has the properties of the model of ENU then – using the relations (8)-(13) – we can determine all its global mutually linearly bound parameters with the accuracy of the estimated parameter (26).

From the relations (8), (12) and (26) result

*the present estimated gauge factor of our ENU (Universe)*

$$a_{pres} \sim (1.1-1.7) \times 10^{26} \text{ m}, \tag{27}$$

and

*the present estimated mass density of our ENU*

$$\rho_{pres} \sim (0.55-1.2) \times 10^{-26} \text{ kg m}^{-3}. \tag{28}$$

According to the present astronomical observations, at present time our observed expansive and isotropic relativistic Universe is anisotropic, i.e. the gravitational forces manifested in it only to distances approximately 100–200 Mpc, which represents approximately 1.8–5.4% of its present estimated gauge factor $a_{pres}$ (27). In larger distances it is isotropic.

The mass density of *the hierarchical gravitationally-bound rotational systems* (HGRS) $\rho_{HGRS}$ decreases with the increase in its dimensions. The stellar systems, the star clusters, the galaxies and the galaxy clusters have the mass density $\rho_{HGRS}$ substantially greater than is the present estimated mass density of Universe $\rho_{pres}$ (28). The average supercluster has the mass $m_{sup} \sim 10^{17} M_{sol} \sim 2 \times 10^{47}$ kg (where $M_{sol}$ is the solar mass), the diameter $d_{sup} \sim 100$ Mpc $\sim 3 \times 10^{24}$ m, and the mass density $\rho_{sup} \sim 1.3 \times 10^{-26}$ kg m$^{-3}$. This means that $\rho_{sup}$ is only a little greater than the present estimated mass density of Universe $\rho_{pres}$ (28) and the escape velocity from the supercluster $v_{2\text{-}sup}$ is only a little greater than is the present velocity of Universe expansion with a given dimension $v_r$.

In our observed Universe – with the model properties of the ENU and the estimated gauge factor $a_{pres}$ (27) – only the HGRS with the mass density $\rho_{HGRS} > \rho_{pres}$ (28) can exist. Therefore, at present time the HGRS with greater dimensions than the superclusters have, cannot exist (Skalský and Súkeník 1993).

The expansive and isotropic relativistic universe – which is in the Newtonian approximation described by the ENU model relations (8)-(13) – has the special-relativistic properties.

In the expansive and isotropic relativistic universe with the special-relativistic properties for the space-time relations are valid *the Lorentz transformations*:

$$x' = \frac{x-vt}{\sqrt{1-\frac{v^2}{c^2}}}, \tag{29}$$

$$y' = y, \tag{30}$$

$$z' = z, \tag{31}$$

$$t' = \frac{t-\frac{v}{c^2}x}{\sqrt{1-\frac{v^2}{c^2}}}, \tag{32}$$

where $x'$, $y'$, $z'$ are the space co-ordinates and $t'$ is the time in the inertial system, which move relative to the observer at the velocity $v$; $x$, $y$, $z$ are the space co-ordinates and $t$ is the time in own observer inertial system.

For the moving special-relativistic mass $m'$ and for the rest mass $m$ of matter object in the inertial co-ordinate system, which move relative to the observer at the velocity $v$, is valid the relation:

$$m' = \frac{m}{\sqrt{1-\frac{v^2}{c^2}}}. \tag{33}$$

The ENU during the whole expansive evolution expands at the boundary velocity of signal propagation $c$. The velocity $c$ does not depend on the velocity of moving objects. Therefore, all observers in the ENU are in its "centre" and – according to the relations (8)-(13) – in the Newtonian approximation they "observe" it as an expanding Euclidean sphere, which at the distance $r = a$ from they expands at the constant velocity $v = c$.

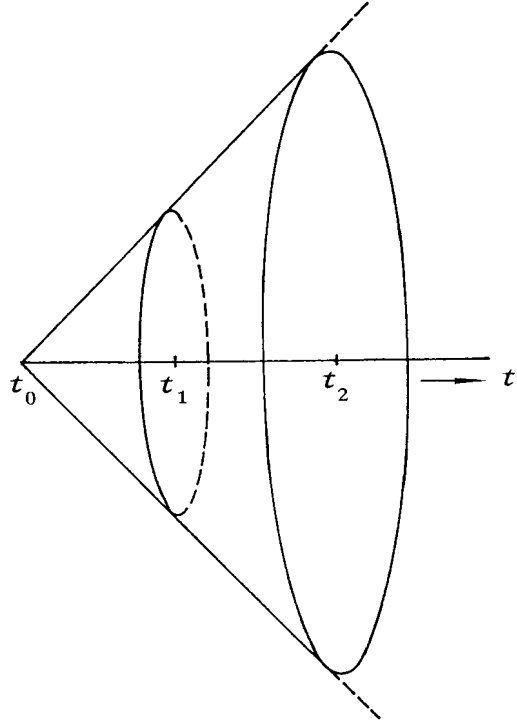

**Figure 1.** The 3-dimensional non-relativistic (Newtonian-Euclidean) projection of the evolution of the 4-dimensional expansive and isotropic relativistic universe.

If we separate the time component from spatial ones, we may show the evolution of the ENU in the form of a time cone. (See Figure 1.)

The ellipses in Figure 1 represent the 2-dimensional non-relativistic (Newtonian-Euclidean) projections of the 3-dimensional space of the expansive and isotropic relativistic universe in the cosmological times $t_1$, $t_2$, ...

Optically we can observe the objects only in the matter era, when the matter separated from the radiation and the expansive and isotropic relativistic universe became transparent for photons. In our observed Universe at present time it represents approximately 99,995% of its present gauge factor $a_{pres}$.

According to the Lorentz transformation of time (32), in the expansive and isotropic relativistic universe the relativistic dilatation of time flow is obtained and at the boundary velocity $v = c$ the time flow is stopped. Therefore, the expansive and isotropic relativistic universe has a pseudo-Euclidean geometry and is closed in space-time manner.

The relation (32) gives the relation for relativistic dilatation of time:

$$t' = \frac{t}{\sqrt{1 - \frac{v^2}{c^2}}}. \tag{34}$$

The relativistic dilatation of time, determined by the relation (34), is projected in Figure 2.

For the obviousness, in Figure 3 there is projected the relativistic dilatation of time in the expansive and isotropic relativistic universe projected into the 2-dimensional projection of the Friedmannian ENU at any cosmological time $t$.

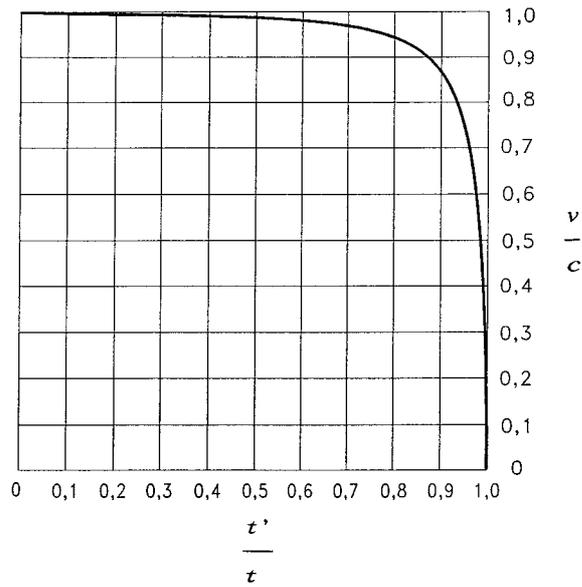

**Figure 2.** The relativistic dilatation of time.

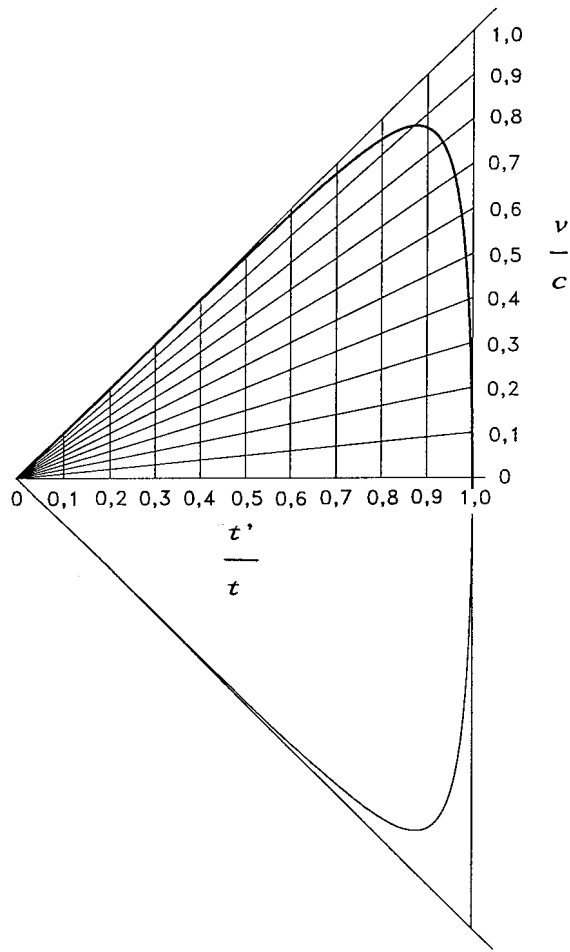

**Figure 3.** The relativistic dilatation of time in the expansive and isotropic relativistic universe projected into the 2-dimensional Friedmannian model of ENU at any cosmological time $t$.

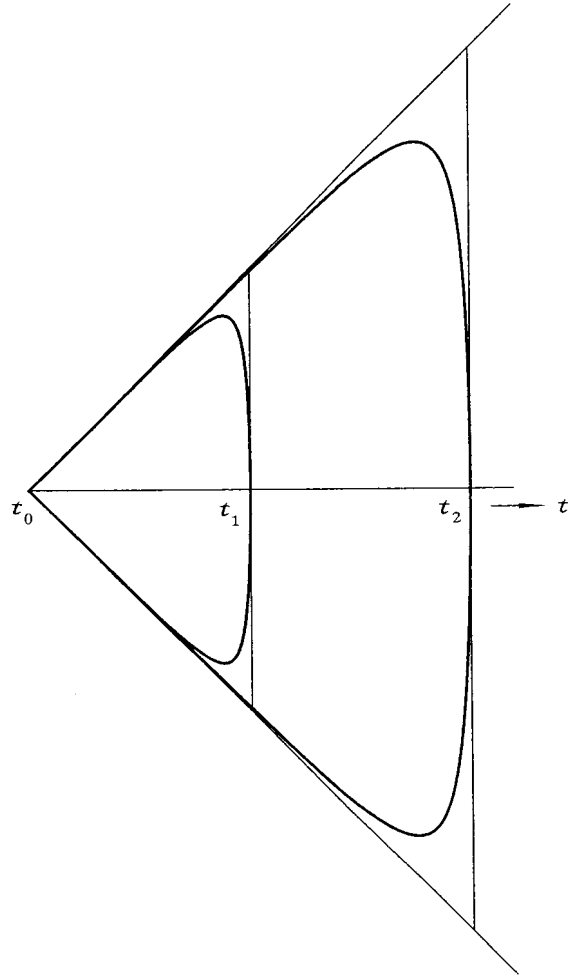

**Figure 4.** The 2-dimensional projection of the 4-dimensional expansive and isotropic relativistic universe projected into the 2-dimensional projection of the Friedmannian model of ENU.

The time cone in Figure 4 represents *the horizon of (all) events*. The curves for observers in points $t_1, t_2, ...$ connect relatively simultaneous events, therefore, represent *the optical horizon (the horizon of visibility, the horizon of particles)* (Rindler 1956). The points on the inner side of curves represent *the past events*. The points between the horizon of events (the time cone) and the external side of the curves represent *the future events*.

According to the relation (33), the moving objects with the rest mass $m$ with the boundary velocity $v = c$ would have an infinite special-relativistic moving mass $m'$. Therefore, the objects with the rest mass principally cannot move at the boundary velocity.

In Figure 4 we can see that the expansive and isotropic relativistic universe (projected into the Friedmannian flat ENU) is relativistically pseudo-Euclidean closed in space-time manner and during the whole expansive evolution in the utmost distance towards to each observer $r = a$, i.e. in the point $t' \equiv t_0$, it expands at the boundary (limit) velocity.

According to the quantum mechanics, the matter objects in the expansive universe can originate at the time $t > t_P$, where $t_P$ is the Planck time. Therefore, although the expansive and isotropic relativistic universe in the Newtonian approximation as a whole expands at the limit velocity $v = c$, the objects with rest mass relative to other objects with rest mass (i.e. relative to other relativistic co-ordinate systems) in it expands at the velocities $v < c$.

We can see in Figure 4 that the relativistic dilatation of time, the relativistic stopping of time at the limit velocity, and the increase of relativistic mass of expanding objects with the rest mass in the expansive and isotropic relativistic universe with the Minkovski pseudo-Euclidean geometry may occur without any paradoxes.

In the expansive and isotropic relativistic universe the length of moving matter objects contracts relativistically. The relation (29) gives the Lorentz-Fitzgerald relation for the special-relativistic contraction of the moving matter objects:

$$L' = L\sqrt{1 - \frac{v^2}{c^2}}, \qquad (35)$$

where $L'$ is the radial length of matter object moving at radial velocity $v$ and $L$ is their proper radial length.

The values of Lorentz-Fitzgerald contraction of moving matter objects $L'$ (35) at some selected velocities are shown in Table I.

The Lorentz-Fitzgerald contraction of the unit spherical moving matter object is shown in Figure 5.

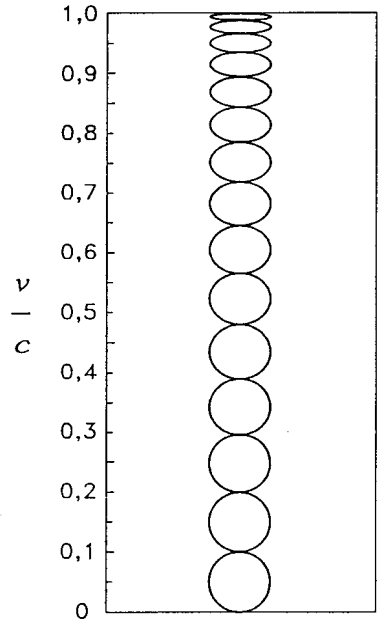

**Figure 5.** The 2-dimensional projection of the relativistic contraction radial length of unit matter sphere, expanding at some chosen velocities.

The Doppler red shift $z_D$ is determined by the relations:

$$z_D = \sqrt{\frac{c+v}{c-v}} - 1 \equiv \frac{1 + \frac{v}{c}}{\sqrt{1 - \frac{v^2}{c^2}}} - 1, \qquad (36)$$

where $v$ is the velocity of radiation source.

The values of Doppler red shift $z_D$, determined by the relations (36), at some selected velocities are shown in Table I.

**Table I**

The values of the Doppler red shift $z_D$, the Lorentz-Fitzgerald relativistic contraction of radial length $L'$ and the seeming dilatation of altitude $A'$ (or the seeming dilatation of breadth $B'$) of the expanding objects at any chosen velocities $v$.

| $v/c$ | $z_D$ | $L'/L$ (%) | $A'/A$ ($B'/B$) (%) |
|---|---|---|---|
| 0 | 0 | 100 | 100 |
| 0.1 | 0.106 | 99.499 | 100.504 |
| 0.2 | 0.225 | 97.980 | 102.062 |
| 0.3 | 0.363 | 95.394 | 104.828 |
| 0.4 | 0.528 | 91.652 | 109.109 |
| 0.5 | 0.732 | 86.603 | 115.470 |
| 0.6 | 1.000 | 80.000 | 125.000 |
| 0.7 | 1.380 | 71.414 | 140.028 |
| 0.8 | 2.000 | 60.000 | 166.667 |
| 0.9 | 3.359 | 43.589 | 229.416 |
| 0.91 | 3.607 | 41.461 | 241.192 |
| 0.92 | 3.899 | 39.192 | 255.155 |
| 0.93 | 4.251 | 36.756 | 272.065 |
| 0.94 | 4.686 | 34.117 | 293.105 |
| 0.95 | 5.245 | 31.225 | 320.256 |
| 0.96 | 6.000 | 28.000 | 357.143 |
| 0.97 | 7.103 | 24.310 | 411.345 |
| 0.98 | 8.950 | 19.900 | 502.519 |
| 0.99 | 13.107 | 14.107 | 708.881 |
| 0.991 | 13.874 | 13.386 | 747.039 |
| 0.992 | 14.780 | 12.624 | 792.155 |
| 0.993 | 15.873 | 11.811 | 846.637 |
| 0.994 | 17.230 | 10.938 | 914.243 |
| 0.995 | 18.975 | 9.987 | 1001.252 |
| 0.996 | 21.338 | 8.935 | 1119.154 |
| 0.997 | 24.801 | 7.740 | 1291.964 |
| 0.998 | 30.607 | 6.321 | 1581.930 |
| 0.999 | 43.710 | 4.471 | 2236.627 |
| 0.9991 | 46.130 | 4.242 | 2357.553 |
| 0.9992 | 48.990 | 3.999 | 2500.500 |
| 0.9993 | 52.443 | 3.741 | 2673.080 |
| 0.9994 | 56.726 | 3.464 | 2887.184 |
| 0.9995 | 62.238 | 3.162 | 3162.673 |
| 0.9996 | 69.704 | 2.828 | 3535.888 |
| 0.9997 | 80.644 | 2.449 | 4082.789 |
| 0.9998 | 98.995 | 2.000 | 5000.250 |
| 0.9999 | 140.418 | 1.414 | 7071.245 |
| 0.99991 | 148.068 | 1.342 | 7453.728 |
| 0.99992 | 157.111 | 1.265 | 7905.852 |
| 0.99993 | 168.028 | 1.183 | 8451.690 |
| 0.99994 | 181.571 | 1.095 | 9128.846 |
| 0.99995 | 198.997 | 1.000 | 10000.125 |

The ENU during the whole expansive evolution expand at the velocity $v = c$, therefore, the 4-dimensional expansive and isotropic special-relativistic universe with pseudo-Euclidean geometry is in the Newtonian approximation (according to the relations (8)-(13)) observed as projected into the 3-dimensional Euclidean expanding sphere. In its utmost distance we observe our early Universe and at the distance $r = a_{pres} - l_P$ (where $l_P$ is the Planck length) is projected our relativistic Universe of the Planck dimensions, which we can observe – at present time only theoretically – by the hypothetical gravitational observation techniques as the most remote and simultaneously the earliest area of our Universe.

For the altitude and the breadth of observed objects in the expansive and isotropic relativistic universe (with the special-relativistic properties) are valid the Lorentz transformations (30) and (31). However, due to the projection of the 4-dimensional pseudo-Euclidean expansive and isotropic special-relativistic universe, closed in space-time manner, into the 3-dimensional Euclidean sphere, expanding at the limit velocity $v = c$, exist – non-considered till now – a seeming dilatation of altitude and breadth of the expanding remote objects.

The seeming dilatation of altitude and breadth of expanding remote objects of the expansive and isotropic relativistic universe in the Newtonian projection (approximation) is determined by the relativistic dilatation of angular dimensions of altitude and breadth of expanding remote objects at the optical horizon, which is the result of relativistic dilatation of time (32). Therefore, in the expansive and isotropic relativistic universe projected in the ENU model for the seeming dilatation of altitude $A'$ and for the seeming dilatation breadth $B'$ of observed remote expanding objects – projected into the 3-dimensional Euclidean sphere which expands at the radial limit velocity $v = c$ from the observer – are valid the relations (Skalský 1994):

$$A' = \frac{A}{\sqrt{1 - \frac{v^2}{c^2}}}, \tag{37}$$

$$B' = \frac{B}{\sqrt{1 - \frac{v^2}{c^2}}}, \tag{38}$$

where $A$ is its rest altitude and $B$ is its rest breadth.

The substance of seeming dilatation of breadth $B'$ in the expansive and isotropic relativistic universe projected into expanding Euclidean sphere is shown in Figure 6.

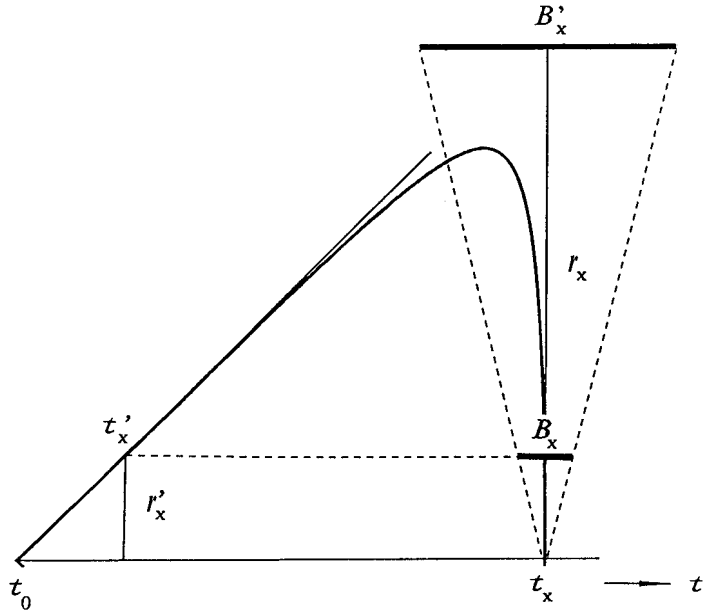

**Figure 6.** The seeming dilatation of breadth of expanding remote object in the expansive and isotropic relativistic universe projected into the Friedmannian model of ENU.

In Figure 6 we see that in the expansive and isotropic relativistic universe projected into an expanding Euclidean sphere the remote object expanding at the radial velocity $v_x$ (in the result the relativistic dilatation of time (32), or (34)) is in the point $t'_x$, i.e. at the radial distance $r'_x$ from the observer in the point $t_x$. Its breadth $B_x$ in the 4-dimensional expansive and isotropic special-relativistic universe with the pseudo-Euclidean geometry is observed (according to the Friedmannian model relations (8)–(13)) projected into the radial distance $r_x$, which corresponds to the velocity $v_x$ in the 3-dimensional Euclidean sphere expanding at the limit velocity $v = c$ from observer in the point $t_x$. The result of this projection is the seeming dilatation of breadth $B'_x$.

The values of seeming dilatation of altitude $A'$ (37) and simultaneously the values of seeming dilatation breadth $B'$ (38) of the expanding remote objects in the expansive and isotropic relativistic universe projected into the model of ENU at any chosen velocities are shown in Table I.

The Lorentz-Fitzgerald contraction of the radial length $L'$ (35) with the simultaneous seeming dilatation of breadth $B'$ (38) of unit spherical expanding remote object in the expansive and isotropic relativistic universe projected into the model of ENU at any chosen velocities is shown in Figure 7.

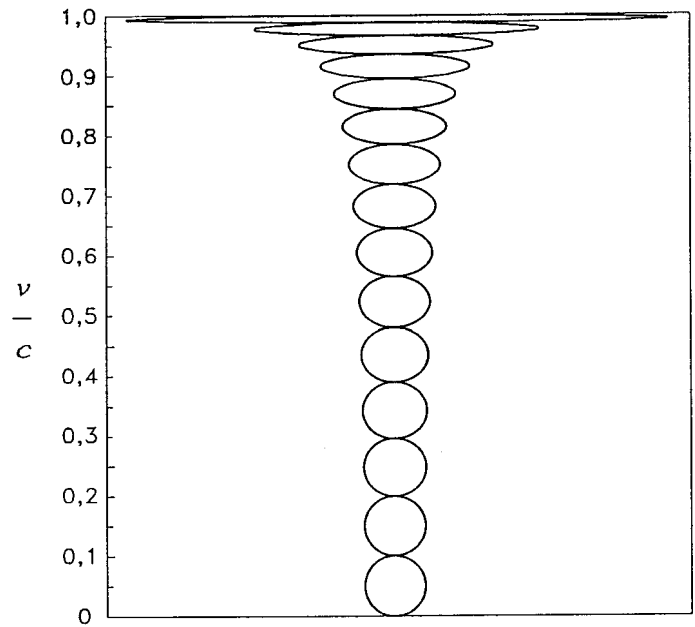

**Figure 7.** The 2-dimensional projection of the relativistic contraction of radial length and simultaneously that of the seeming dilatation breadth of unit matter sphere in the ENU, expanding at any chosen velocities.

From the relations $A'$ (37) and $B'$ (38) it results that in the expansive and isotropic special-relativistic universe with the pseudo-Euclidean geometry, projected into the model of ENU, the angular dimensions of the expanding objects decreases only up to the distance at which they expand at the radial velocity (Skalský 1994)

$$v_r = 2.350 \times 10^8 \text{ ms}^{-1} \tag{39}$$

and have the Doppler red shift

$$z_D = 1.873, \tag{40}$$

i.e. only to the inverse distance

$$r_i = 78.388\% \, a, \tag{41}$$

where $a$ is the gauge factor of ENU.

At the distances $r > r_i$ the angular dimensions of the observed objects increases (Skalský 1994). (See Figure 7.)

Due to the Lorentz-Fitzgerald contraction $L'$ (35), the seeming dilatation of altitude $A'$ (37) and the seeming dilatation of breadth $B'$ (38), the very distant expanding objects in the expansive and isotropic relativistic universe, projected into the model of ENU, are observed as *the big discs*, *the big circles* or *"the big flat chimeras"* (Skalský 1994).

If our observed expansive and isotropic relativistic Universe have the model mass-space-time properties of the ENU then the Lorentz-Fitzgerald contraction of length $L'$ (35) with the simultaneous hypothetical seeming dilatation of altitude $A'$ (37) and the hypothetical seeming dilatation of breadth $B'$ (38) of very distant expanding objects should be observable by the present advanced observation technology, or by the observational technology which we will have at our disposal in the nearest future.

On January 15, 1996 *the Space Telescope Science Institute* (STScI) published the famous *Hubble Deep Field* with dilated dimensions of the very distant cosmic objects (the galaxies) (STScI 1996). Late there were discovered even more distant cosmic objects with dilated dimensions.

In the present cosmological literature the observed dilatation of dimensions of very distant cosmic objects is interpreted solely as a result of gravitational lenses. However, we assume that the hypothetical seeming dilatation of altitude $A'$ (37) and the hypothetical seeming dilatation of breadth $B'$ (38) participate on the observed dilatation of the dimensions of the very distant objects with a significant proportion.

Whether on the dilatation of dimensions of observed very distant cosmic objects do participate the hypothetical seeming dilatation of altitude $A'$ (37) and the hypothetical seeming dilatation of breadth $B'$ (38), too, may be decided by observation, because between dilatation of remote cosmic objects caused by the gravitational lenses and by the hypothetical special-relativistic seeming dilatation of altitude and breadth there exist these significant differences:

1. The dilatation of cosmic objects caused by the gravitational lenses is demonstrated only locally (near the object, which has the function of the gravitational lens). The hypothetical special-relativistic seeming dilatation of altitude $A'$ (37) and the hypothetical special-relativistic seeming dilatation of breadth $B'$ (38) manifest themselves globally at the extent of the whole horizon of universe.

2. The gravitational lenses cause equal dilatation of all dimensions (i.e. altitude, breadth and length). The hypothetical special-relativistic seeming dilatation manifest itself only at the altitude and the breadth (their pendant is the special-relativistic contraction of length $L'$ (35)). (See Figure 7.)

3. The gravitational lenses cause the equal dilatation of all dimensions of observed objects at arbitrary distance from observer. The hypothetical special-relativistic seeming dilatation of altitude $A'$ (37) and the hypothetical special-relativistic seeming dilatation of breadth $B'$ (38) at distance steeply grow. (See Figure 7.)

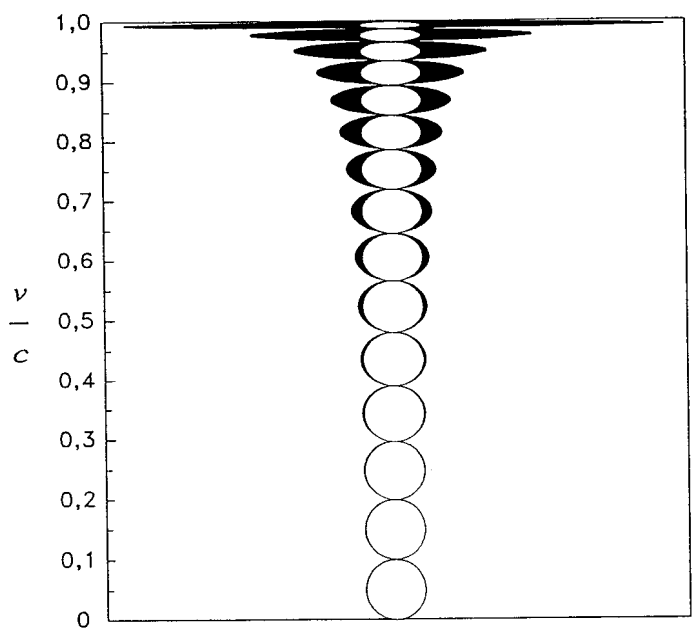

**Figure 8.**   The projection of difference between how the remote expanding spherical matter objects should look like according to the model of ENU and how they should look according to all the other models of expansive relativistic universe. The difference is projected in black.

The difference between this how the remote expanding spherical matter objects should look like according to the model of ENU and according to the other models of the expansive relativistic universe is shown in Figure 8.

We can conclude:

The model of ENU – determined by the Friedmannian equations (1a), (1b) and (1c) with $k = 0$, $\Lambda = 0$ and $K = -1/3$ – has such mass-space-time properties, which make it unambiguously different from the other models of universe. The manifestations of these properties are such distinctive that – under the assumption that our expansive and isotropic relativistic Universe in the Newtonian approximation has the properties of the ENU model – they should be observable by the high observation technology (Skalský 1994, 1997).

## 5. The analysis of the Friedmannian models of universe

The Friedmannian models of isotropic and homogeneous universe, determined by the Friedmannian equations (1a), (1b) and (1c) with $k = +1, 0, -1$; $\Lambda > 0, = 0, < 0$; and $K > 0, = 0, < 0$, describe infinite number of models of static and dynamic isotropic relativistic universe.

The Friedmannian models of static universe, described by the Friedmannian equations (1a), (1b) and (1c), are the generalisation of three groups of static universe models:

1. - The Einstein static model of spherical (Riemannian) universe with $k = +1$, $\rho > 0$ and $\Lambda > 0$ (Einstein 1917).
   - The Friedmann static model of hyperbolic (Lobachevskian) universe with $k = -1$, $\rho < 0$ and $\Lambda < 0$ (Friedmann 1924).
2. - The de Sitter static model of elliptical (Newcombian) universe with $k = +1$, $\rho = 0$ and $\Lambda > 0$ (de Sitter 1917).
   - The Friedmann static model of hyperbolic (Lobachevskian) universe with $k = -1$, $\rho = 0$ and $\Lambda < 0$ (Friedmann 1924).
3. - The de Sitter static model of flat (Euclidean) universe with $k = 0$, $\rho = 0$ and $\Lambda = 0$ (de Sitter 1917).

In 1930 A. S. Eddington draw attention to the fact that *the Einstein first* (static isotropic and homogeneous) *relativistic model of universe* (Einstein 1917) is extremely non-stable, because an arbitrary small fluctuation would change it into the non-static one (dynamic) (Eddington 1930).

Einstein in 1931 accepted the Eddington argument and abandoned his static model of universe (Einstein 1931).

The Eddington's argument is valid on the Friedmann static model of universe with the non-zero (negative) mass density $\rho$ (Friedmann 1924), and on the next possible equally balanced two-part variants (with the non-zero mass density $\rho$ and with the non-zero value of cosmological number $\Lambda$), described by the Friedmannian equations (1a), (1b) and (1c), too.

The second and third groups of the Friedmannian models of static universe have only formal meaning.

The analysis of the Friedmannian models of dynamic universe we begin by the analysis of the Friedmannian models of the flat expansive isotropic and homogeneous universe.

This way was chosen intentionally, because these models are most transparent. There is no doubt about how they are projected, therefore, we can simply control their model properties.

The Friedmannian models of flat expansive isotropic and homogeneous universe *de facto* represent the Newtonian-Euclidean expanding homogeneous matter sphere, determined by the Friedmannian equations (1a), (1b) and (1c) with $k = 0$, $\Lambda = 0$ and $K > 0, = 0, < 0$.

In the Newtonian theory of gravitation the matter objects are considered as the point masses in the Euclidean space, which are equivalent to the equal spherically symmetrically display of mass.

The Newtonian-Euclidean homogeneous matter sphere is unambiguously determined by two parameters:

1. the point mass $m$,
2. the radius $r$.

This means that the individual Friedmannian models of flat universe are in the finite result manifested by a different proportion ratio of these two above parameters. Therefore, the gauge factor of Friedmannian flat expansive homogeneous universe $a$ can be interpreted by the relation:

$$a = vt, \tag{42}$$

where $v$ is the velocity of universe expansion and $t$ is the cosmological time.

The relativistic and quantum-mechanics conditions – expressed by the relations (2), (3), (4), (5) and (6) – are fulfilled only by the Friedmannian model of flat expansive isotropic and homogeneous universe (ENU), determined by the Friedmannian equations (1a), (1b) and (1c) with $k = 0$, $\Lambda = 0$ and $K = -1/3$.

In the ENU are valid the relations $a = r_g = ct$, therefore, during the whole expansive evolution it expands at the velocity $v = c$.

In all other Friedmannian models of flat expansive universe, determined by the Friedmannian equations (1a), (1b) and (1c) with $k = 0$, $\Lambda = 0$ and $K \neq -1/3$, are valid the relations $a \neq r_g$ and they expand at the velocities $v \neq c$. This means that either in it is valid the relation $a < r_g$, what contradicts *the Schwarzschild first exact solution of the Einstein equations of gravitational field* (Schwarzschild 1916) and *"no hair theorem"* (Carter 1971; Hawking 1972; Robinson 1975) or they expand at the velocities $v > c$, what contradicts the one of fundamental principles of the theory of relativity (Einstein 1905, 1915, 1916, 1954).

From infinite number of the Friedmannian models of flat expansive isotropic and homogeneous universe we demonstrate it on the three flat variants so-called *the standard model of universe*, determined by the Friedmannian equations (1a) and (1b) with $k = 0$ and $\Lambda = 0$ and three state equations:

*the boundary hard state equation*

$$p = \varepsilon, \tag{43}$$

*the ultra-relativistic state equation*

$$p = \frac{1}{3}\varepsilon, \tag{44}$$

*the dust state equation*

$$p = 0. \tag{45}$$

For the gauge factor $a$ and the cosmological time $t$ of the flat variant of the standard universe model, which is determined by the Friedmann equations (1a) and (1b) with $k = 0$ and $\Lambda = 0$ and the boundary hard state equation (43), is valid the relation (Monin *et al.* 1989):

$$a = \sqrt[3]{3ca_0^2 t}, \tag{46}$$

where $a_0$ is the chosen scale.

The relation (46) with $a_0 = l_P$ (5) and $t = t_P$ (6) gives the value of gauge factor

$$a = 2.330\ 74(69) \times 10^{-35}\ \text{m} = 144.225\ \%\ l_P \tag{47}$$

and – according to the relations (42) and (47) – it expands at the velocity

$$v = 4.323\ 75(53) \times 10^8\ \text{m s}^{-1} = 144.225\ \%\ c. \tag{48}$$

If in the relation (46) we put $a = a_0 = l_P$ (5) than the cosmological time

$$t = 1.796\ 85(37) \times 10^{-44}\ \text{s} = 33.333\ \%\ t_P. \tag{49}$$

For the gauge factor $a$ of the flat variant of the standard universe model, which is determined by the Friedmann equations (1a) and (1b) with $k = 0$ and $\Lambda = 0$ and the ultra-relativistic state equation (44), is valid the relation (Monin *et al.* 1989):

$$a = \sqrt{2ca_0 t}, \tag{50}$$

i.e. the gauge factor $a$, determined by the relation (50) with $a_0 = l_P$ (5) and $t = t_P$ (6), has the value:

$$a = 2.285\ 43(93) \times 10^{-35}\ \text{m} = 141.421\ \%\ l_P \tag{51}$$

and – according to the relations (42) and (51) – it expands at the velocity

$$v = 4.239\ 70(55) \times 10^8\ \text{m s}^{-1} = 141.421\ \%\ c. \tag{52}$$

If in the relation (50) we put $a = a_0 = l_P$ (5) than the cosmological time

$$t = 2.695\ 28(06) \times 10^{-44}\ \text{s} = 50\ \%\ t_P. \tag{53}$$

For the gauge factor $a$ of the flat variant of the standard universe model, which is determined by the Friedmann equations (1a) and (1b) with $k = 0$ and $\Lambda = 0$ and the dust state equation (45), is valid the relation (Monin *et al.* 1989):

$$a = \sqrt[3]{\frac{9a_0 c^2 t^2}{4}}, \tag{54}$$

i.e. the gauge factor $a$, determined by the relation (54) with $a_0 = l_P$ (5) and $t = t_P$ (6), has the value:

$$a = 2.117\ 62(41) \times 10^{-35}\ \text{m} = 131.037\ \%\ l_P \tag{55}$$

and – according to the relations (42) and (55) – it expands at the velocity

$$v = 3.928\ 39(25) \times 10^8\ \text{m s}^{-1} = 131.037\ \%\ c. \tag{56}$$

If in the relation (54) we put $a = a_0 = l_P$ (5) than the cosmological time

$$t = 3.593\ 70(75) \times 10^{-44}\ \text{s} = 66.667\ \%\ t_P. \tag{57}$$

Therefore, we can state:

The Friedmannian models of the flat expansive isotropic and homogeneous universe, determined by the Friedmannian equations (1a), (1b) and (1c) with $k = 0$, $\Lambda = 0$ and $K \neq -1/3$ (including the flat variants of so-called standard model of universe, determined by the Friedmannian equations (1a), (1b) and (1c) with $k = 0$, $\Lambda = 0$ and $K = 1, 1/3, 0$), contradict the general theory of relativity and the quantum mechanics (Skalský 2000b, c).

More detailed analysis of the problem of velocity expansion of the standard model of universe can be found in the papers *The Planck quantum hypothesis and the Friedmannian models of flat universe* (Skalský 2000b) and *The gauge factor increase and the hypothetical emerging of matter objects on the horizon in the standard model of universe* (Skalský 2000c).

Now we verify the possibility of existence of the Friedmannian models which are determined by the Friedmannian equations (1a), (1b) and (1c) with $K = -1/3$, $\Lambda = 0$ and $k = +1, -1$.

All Friedmannian models of expansive universe are expressed in the Newtonian approximation, therefore, they are differ by a different proportion of the point mass $m$ and the radius of expansive matter sphere $r$. This means that the index of curvature $k$ in the Friedmannian models of universe with the Newtonian-Euclidean properties would express the overcritical, critical or subcritical mass density $\rho$ in the only possible manner: by the different proportion of the point mass $m$ and the gauge factor $a$ (i.e. the radius $r$) in the Euclidean homogeneous matter sphere.

For obviousness this hypothetical possibility is shown in Figure 9.

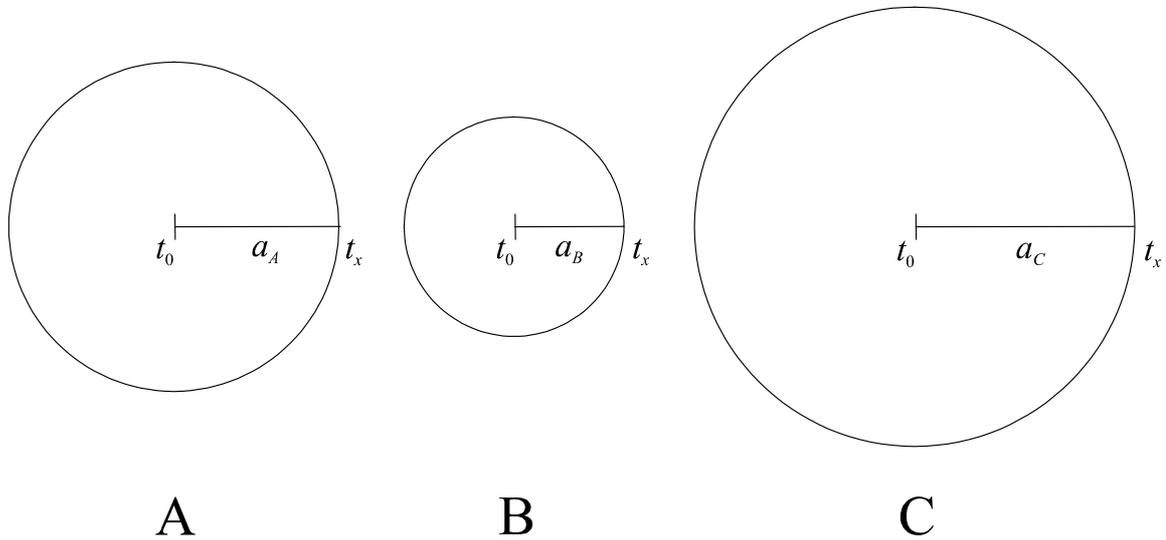

**Figure 9.** The 2-dimensional projection of the Newtonian models of the hypothetical expansive and isotropic relativistic universes.

The circles A, B and C in Figure 9 represent the two-dimensional projection of the 3-dimensional Euclidean homogeneous matter sphere with equal mass $m$, but with different radius $r$ (i.e. with different mass density $\rho$), which represent the Newtonian projection of the hypothetical expansive and isotropic relativistic universes at any cosmological time $t_x$.

Assume that the mass density $\rho$ of the Newtonian model of expansive and homogeneous universe A in Figure 9 is critical. This means that the model of universe A expands at the exact escape velocity $v_2 = c$ and its gauge factor $a_A = r_g$.

The Newtonian model of expansive and homogeneous universe B in Figure 9 has the overcritical mass density $\rho$, i.e. its gauge factor $a_B < r_g$. This means that the model of universe B with the overcritical mass density $\rho$, contradict the general theory of relativity, because the relativistic universe in the Newtonian approximation principally cannot have the gauge factor $a < r_g$!

The Newtonian model of expansive and homogeneous universe C in Figure 9 has the subcritical mass density $\rho$. This means that its gauge factor $a_C > a_A$, i.e. the universe C, expands at the velocity $v > c$. Hence the universe C with the subcritical mass density $\rho$, contradicts the general theory of relativity, because the expansive and isotropic relativistic universe in the Newtonian approximation principally cannot expand at the velocity $v > c$!

At the end we verify the possibility of the existence of models which are determined by the Friedmannian equations (1a), (1b) and (1c) with $K = -1/3$, $k = 0$ and $\Lambda \neq 0$.

In the Friedmannian models of expansive universe, which *de facto* represent the expanding Newtonian-Euclidean homogeneous matter sphere, the value of cosmological member $\Lambda$ (equally as the different value of the index of curvature $k$) can manifest itself only by change of the proportion of the point mass $m$ and the gauge factor $a$ (i.e. the radius $r$). This means that the values of $\Lambda > 0$ and $\Lambda < 0$ in the Friedmannian models of universe would lead to breaking of the relativistic and quantum-mechanical restrictive conditions which are expressed in the relations (2), (3), (4), (5) and (6).

## 6. Conclusions

The relativistic cosmology is constructed on the basis of the general theory of relativity and its special partial solutions: the special theory of relativity, in which we abstract from the gravitational effects, and the Newton theory of gravitation (the classical mechanics), in which we abstract from all relativistic effects.

A. Einstein formulated the final version of the general theory of relativity in 1915 (Einstein 1915). Over half-century the general theory of relativity was verified only in the weak gravitational field, but the crucial verifying of the general theory of relativity in the strong gravitational field was absent. However, the situation was quantitatively and qualitatively changed after the discovery of the binary pulsar PSR 1913+16 in 1974 by R. A. Hulse and J. H. Taylor (the Nobel prizewinners in 1993) (Taylor *et al.* 1979, Hulse 1994, Taylor 1994) and other binary pulsars, such as B1534+12, B2127+11C, and B1855+09 (Wolszczan 1997).

The binary pulsars – due to they extraordinary properties – give the unique possibility for verifying the general theory of relativity in the strong gravitational field with accuracy, which is substantially dependent on a period of observations.

The results of twenty-year observations of the pulsar PSR 1913+16 are consistent with the predictions of the general theory of relativity with preciseness of $10^{-14}$ (Penrose 1997). This means that the general theory of relativity has become one of the most verified scientific theories.

Einstein often stressed the logical completion of the general theory of relativity. For example in the article *What is the theory of relativity?* he wrote about it that an attractive part of this theory is its logical completion. If any of its conclusions appears wrong, it must be rejected; any of its modifications that do not break its structure is impossible (Einstein 1954).

The Einstein argumentation can be reversed, too:

**If any hypothetical supplementary assumption introduced into the applications of the general theory of relativity (or into the applications of its special partial solutions: the special theory of relativity and the classical mechanics), leads to the solution which contradicts its principles or its conclusions, then this hypothetical supplementary assumption is – from stand point of the general theory of relativity – non-correct and non-admissible!**

In this work we show only some from the hypothetical supplementary assumptions introduced into the relativistic cosmology, which contradicts the general theory of relativity. Other ones we analysed in other works, some of which are (or in short time will be) accessible on the Internet, too (Skalský 1999, 2000a, b, c, d, e, f).

The introduction of the hypothetical supplementary assumptions, which contradict the general theory of relativity and the quantum mechanics into the relativistic cosmology, has also deeper gnoseological and formally logical significance, because it means (consciously or non-consciously) violation of *the fundamental formal principles of logical thinking* (Skalský 2000g):

1. *the principle of identity (principium identitatis)*,

2. *the principle of contradiction (principium contradictionis)*,

3. *the principle of third exclusion (principium exclusii tertii)*,

4. *the principle of rational sufficiency (principium rationis sufficientis)*.

If from the Friedmannian equations (1a), (1b) and (1c) we exclude the models with the properties which contradict the relativistic and quantum-mechanical restrictive conditions – expressed by the relations (2), (3), (4), (5) and (6) – we receive the final version of *the equations of universe dynamics* (Skalský 1993, 1997, 2000d):

$$8\pi G a^2 \rho - 3c^2 = 0, \tag{58a}$$

$$8\pi G a^2 p + c^4 = 0, \tag{58b}$$

$$\varepsilon + 3p = 0, \tag{58c}$$

or some of variants from their one-line forms, for example as:

$$\varepsilon = \frac{3c^4}{8\pi G a^2} = -3p. \tag{59}$$

The equations of universe dynamics – which we can express with the finite values of the universe global parameters only in the Newtonian approximation as the parameters of the flat (Euclidean) expansive nondecelerative homogeneous matter sphere – we can express in other forms, too. For example, in the form:

$$a = ct = \frac{2Gm}{c^2}, \tag{60}$$

or in other forms, which express the relations between minimum three parameters of the universe in the Newtonian approximation, which are contained in the equations (8)–(13).

http://xxx.lanl.gov/abs/astro-ph/0003192